\documentclass[sigconf,preprint,nonacm]{acmart}
\makeatletter                   
\def\mdseries@tt{m}             
\makeatother                    

\usepackage[utf8]{inputenc}
\usepackage{graphicx}
\usepackage{tablefootnote}
\usepackage{array}
\usepackage[draft=true]{minted}
 \usepackage{adjustbox}
\usepackage{standalone}
\usepackage{tikz}
\usetikzlibrary{trees,positioning,shapes,shapes.geometric,arrows,fit,matrix,positioning}

\newcommand{\etape}[2]{node (p#1) [etape] {#2}}

\newcommand{\transreceptor}[3]{%
  \path [linepart] (#1.east) -- node [above] {\scriptsize #2} (#3);}

\usepackage{pgfplots}
\usepackage{courier}
\usepackage{listings} 
\lstdefinestyle{sharpc}{language=[Sharp]C}
\definecolor{forestgreen}{rgb}{0.13, 0.54, 0.13}

\usepackage{listings}
\usepackage{color}

\definecolor{dkgreen}{rgb}{0,0.6,0}
\definecolor{gray}{rgb}{0.5,0.5,0.5}
\definecolor{mauve}{rgb}{0.58,0,0.82}

\usepackage{xcolor,listings}

\pgfplotsset{compat=1.11,
    /pgfplots/ybar legend/.style={
    /pgfplots/legend image code/.code={%
       \draw[##1,/tikz/.cd,yshift=-0.25em]
        (0cm,0cm) rectangle (0.5em,0.5em);},
   },
}
\usepackage[framemethod=tikz]{mdframed}

\definecolor{rqcolor}{rgb}{1.0, 0.95, 0.85}
\newmdenv[backgroundcolor=rqcolor,innerlinewidth=0.5pt, roundcorner=4pt,linecolor=rqcolor,innerleftmargin=6pt,
innerrightmargin=6pt,innertopmargin=6pt,innerbottommargin=6pt]{rqbox}

\usetikzlibrary{shadows,arrows.meta,positioning,backgrounds,fit}
\usetikzlibrary{shapes.geometric}
\tikzset{%
  materia/.style={draw, fill=red!20, text width=5.0em, text centered, minimum height=1.5em},
  etape/.style={materia, text width=8em, minimum width=5em, minimum height=3em, rounded corners},
  texto/.style={above, text width=8em, text centered},
  linepart/.style={draw, thick, color=black!50, -LaTeX, dashed},
  line/.style={draw, thick, color=black!50, -LaTeX},
  ur/.style={draw, text centered, minimum height=0.01em},
  back group/.style={fill=white!20,rounded corners, draw=black!50, dashed, inner xsep=15pt, inner ysep=15pt},
}

\title{DeepPERF: A Deep Learning-Based Approach For Improving Software Performance}
\author{Spandan Garg}
\authornote{Corresponding Author}
\email{spgarg@microsoft.com}
\affiliation{%
 \institution{Microsoft}
 \city{Redmond}
 \state{Washington}
 \country{USA}}
 
\author{Roshanak Zilouchian Moghaddam}
\email{rozilouc@microsoft.com}
\affiliation{%
  \institution{Microsoft}
  \city{Redmond}
  \state{Washington}
  \country{USA}
}

\author{Colin B. Clement}
\email{coclemen@microsoft.com}
\affiliation{%
 \institution{Microsoft}
 \city{Redmond}
 \state{Washington}
 \country{USA}}

\author{Neel Sundaresan}
\email{neels@microsoft.com}
\affiliation{%
 \institution{Microsoft}
 \city{Redmond}
 \state{Washington}
 \country{USA}}
 
 \author{Chen Wu}
\email{chen.wu@microsoft.com}
\affiliation{%
 \institution{Microsoft}
 \city{Shanghai}
 \state{}
 \country{China}}
\date{}

\begin{document}

\begin{abstract}

Improving software performance is an important yet challenging part of the software development cycle. Today, the majority of performance inefficiencies are identified and patched by performance experts. Recent advancements in deep learning approaches and the wide-spread availability of open source data creates a great opportunity to automate the identification and patching of performance problems. In this paper, we present DeepPERF, a transformer-based approach to suggest performance improvements for C\# applications. We pretrain DeepPERF on English and Source code corpora and followed by finetuning for the task of generating performance improvement patches for C\# applications. Our evaluation shows that our model can generate the same performance improvement suggestion as the developer fix in $\sim$53\% of the cases, getting $\sim$34\% of them verbatim in our expert-verified dataset of performance changes made by C\# developers. Additionally, we evaluate DeepPERF on 50 open source C\# repositories on GitHub using both benchmark and unit tests and find that our model is able to suggest valid performance improvements that can improve both CPU usage and Memory allocations. So far we've submitted 19 pull-requests with 28 different performance optimizations and 11 of these PRs have been approved by the project owners. 

\end{abstract}
\maketitle
\thispagestyle{empty}
\pagestyle{plain}

\section{Introduction}
Performance bugs are usually non-functional bugs that can cause poor user experience, reduced throughput, increased latency, and wasted resources. Performance bugs may not cause system failure and may depend on user input, therefore detecting them can be challenging~\cite{perfscope, catchmeifyoucan}. They also tend to be harder to fix than non-performance bugs \cite{nistor2013, song2014oopsla}. As a result, better tool support is needed for fixing performance bugs.  

In recent years, a variety of performance bug detection approaches have emerged to help developers identify performance issues. However, a majority of existing performance bug detection approaches focus on specific types of performance problems. For instance, prior work investigated the detection of inefficient loops \cite{linhai2017icseloops, nistor2013, xiao2013issta}, database related performance issues, low-utility data structures \cite{xu2010pldi}, false sharing specially in multi-threaded code \cite{liu2011oopsal}, etc. Approaches that fix specific performance issues due to repeated computations~\cite{memoization}, software misconfigurations~\cite{misconfigurations}, loop inefficiencies~\cite{caramelnistor}, etc. have also been developed. Many of these approaches rely on expert-written algorithms or pre-defined set of rules to detect and fix performance issues based on patterns in abstract syntax tree, control flow graphs, profiles, etc. Building rule-based analyzers is a non-trivial task as it requires achieving the right balance between precision and recall. Once developed, maintaining these rules can also be costly \cite{bielik2017learning} as it requires continuous effort by performance experts.  

With the recent rise of large transformer models and wide-spread availability of open-source software artifacts, there is an opportunity to learn patterns of performance improvements directly from mined data. Transformer-based approaches have been shown to achieve state-of-the-art performance, not only in various Natural Language Processing (NLP) problems, but also a variety of software engineering tasks such as code-completion~\cite{svyatkovskiy2020intellicode}, documentation generation~\cite{clement2020pymt5}, unit test generation~\cite{Tufano2020UnitTC}, bug detection~\cite{Drain2021DeepDebugFP}, etc. In this paper, we draw inspiration from these techniques, in an attempt to solve the problem of automatically suggesting performance improvements.

We present an approach called DeepDev-Perf that uses a large transformer model to suggest changes at application source code level to improve its performance. We first pretrain our model using masked language modelling (MLM) tasks~\cite{clement2020pymt5} on English text and source code taken from open source repositories on GitHub, followed by finetuning on millions of performance commits made by .NET developers. Through our evaluation, we show that our approach is able to recommend patches to provide a wide-range of performance optimizations in C\# applications, which is not possible through any existing analyzer alone. Most suggested changes involve modifications to high-level constructs like API/Data Structure usages or other algorithmic changes, often spanning multiple methods, which cannot be optimized away automatically by the C\# compiler and could, therefore, lead to slow-downs on the user's side. Further, by suggesting changes to a set of real world repositories and measuring the impact of our suggestions through benchmark tests, we show that our changes provide actual performance gains to these applications. 11 PRs containing our model suggestions have already been accepted by the developers of these projects, showing that our suggestions are considered to be correct and useful by the project owners.


In summary, our work makes the following main contributions:
\begin{itemize} 
    \item We propose a novel transformer-based model called DeepPERF, which finds performance optimization opportunities in a c\# application and automatically generates performance improvement patches. 
    \item We extensively evaluate DeepPERF using a curated dataset of real-world performance improvement changes made by C\# developers to a hold-out set of open source repos on GitHub. Through our empirical evaluation, we demonstrate that Deepdev-PERF is able to generate a wide-variety of performance improvements.  
    \item We show real-world evidence that DeepPERF generates changes that lead to tangible performance improvements to various open source C\# projects on GitHub. We submit PRs containing the suggested changes to these repos, many of which have since been approved showing that our fixes are considered useful by developers.
\end{itemize}

\section{Motivating Examples}
\begin{figure}[h]
\centering
\includegraphics[width=0.38\textwidth]{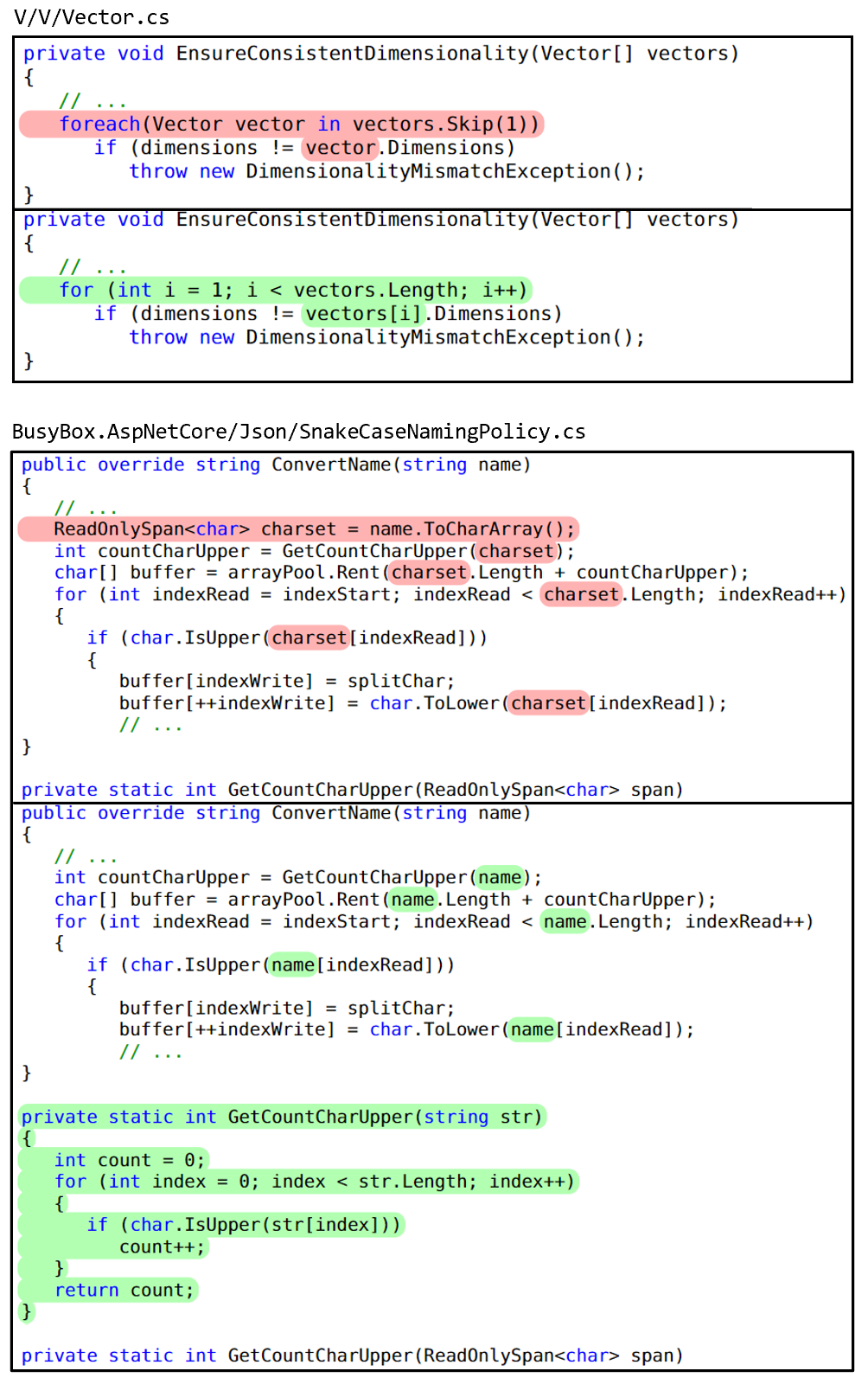}
\caption{Two examples of changes suggested by DeepPERF that were submitted to the corresponding repos as PRs and have since been accepted: (i) the change on top, taken from following PR\protect\footnotemark, suggests a change to unroll a LINQ query into an explicit for-loop. This change results in lower allocations and Gen 0 garbage collection compared to prior code (ii) the change on the bottom is from another PR\protect\footnotemark on a different C\# project. The original code unnecessarily converts a string to a character array to index into the string. Since the array allocation is redundant, DeepPERF suggests a change to remove the allocation in favor of just using the original string instead. It also overloads a user-defined helper method, previously being used to count the number of upper-case characters in the string, to accept a \texttt{string} instead of a \texttt{ReadOnlySpan}.}
\label{motivating_example}
\end{figure}

Figure \ref{motivating_example} shows examples of two suggestions made by DeepPERF to two open-source C\# projects on GitHub. In the first example, the code prior to the change uses LINQ \cite{pialorsi2007introducing}. LINQ expressions have an inherent allocation associated with them. As a result, LINQ usage on the application hot-path often leads to unnecessary allocations, which can cause spikes in garbage collection (GC), depriving the application of CPU resources and reducing its throughput. In the top change, DeepPERF recognizes that the use of LINQ call to skip the first position is unnecessary and it recommends a change to unroll the LINQ query and use an explicit for-loop, which starts indexing from 1. By executing the unit and benchmark tests in this repository, we verified the correctness of the change as well as the performance gain. Looking at the benchmark results, the change achieves a reduction in allocations as well as Gen 0 GC compared to prior code.

The second change shows a case where the code unnecessarily allocates a character array from an input string using the \texttt{ToCharArray} method. The array is then being used to iterate over and index characters at various positions within the string, as well as passed to a user-defined helper function to count the number of uppercase characters within the string. DeepPERF notices that the array allocation is redundant as C\# strings can be indexed directly. Therefore, it removes the redundant allocation and replaces the usages of the array with the original string. It also defines an overload to the helper method that accepts a \texttt{string} instead of a \texttt{ReadOnlySpan} to count the number of uppercase letters in the string. This change results in fewer allocations as well as improved CPU usage. 

Pull-requests containing both of these changes were submitted to the corresponding GitHub repos and have since been approved by their owners.
\footnotetext[1]{\url{https://github.com/CreoOne/V/pull/1}}
\footnotetext[2]{\url{https://github.com/iigoshkin/BusyBox/pull/5}}

\begin{figure*}[h]
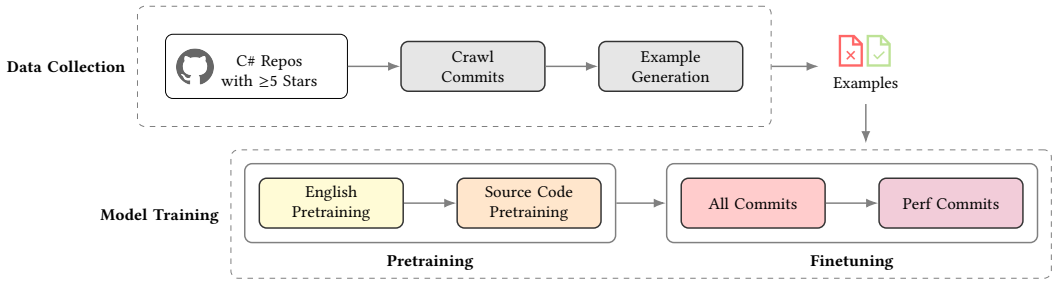

\centering
\includestandalone[width=0.79\textwidth]{other/pipeline}
\caption{Our model data collection and training pipeline. We first crawl the commit history of all 45k C\# repos with $\geq$5 stars on GitHub and generate examples for each modified method with various contextual elements important to performance (using statements, class attributes, caller-callee methods, etc.). For training, we first pretrain BART-large on denoising objectives over English text and source code, followed by a two-step finetuning. In our finetuning step, we first finetune the model on examples generated from all commits, followed by a smaller finetuning step done only on examples from commits where developer included a performance-related keyword ("perf", "performance", "reduce allocation", etc.) in the commit message.}
\label{pipeline}
\end{figure*}

\section{Our Approach}
We describe the details of our proposed model below. Figure \ref{pipeline} shows an overview of our pipeline. We begin by first describing how we take an English-pretrained BART-large model and further pretrain it on Source code. We then describe our data collection and example generation pipeline for finetuning. This is followed by a description of our two-step finetuning process using the examples generated by the example generation step.

\subsection{Pretraining}
For pretraining, we collected 45K GitHub repositories with $\geq$5 stars that were composed primarily of C\# code. We de-duplicate this data on a file-level using a hash function and then pretrain a 406M parameter BART-large model using span-masking objective~\cite{lewis2020bart} on this data set. Span-masking corrupts code by replacing random spans of input tokens with a \texttt{<MASK>} token, and the model is trained to reconstruct the original code by predicting the tokens replaced by the mask tokens. Such pretraining has been shown to significantly improve model performance for a various downstream NLP tasks, including many software engineering tasks~\cite{clement2020pymt5}. 


\subsection{Data Collection}
Below we present the details of our data collection and example generation steps. For our data, we collect $\sim$45k repositories with $\geq$5 stars on Github whose primary language is C\# and had a commit within the last 5 years. 


\subsubsection{Crawling Commits} After cloning these projects, we crawl the \emph{main} branch's commit history. This history involves a commit message and a diff representing the difference between current and previous version of the changed files. This yields $\sim$11M commits and their corresponding commit messages. 

\subsubsection{Identifying Performance-related Commits}
For each commit, we look for performance related keywords ("perf", "performance", "reduce allocation", etc.) in the commit message to determine if it is performance related. Table \ref{perf_non_perf} shows the number of commits and examples that come from performance-related commits. 

\subsubsection{Generating Code Transformation Pairs}
\label{generating_examples}
To generate the code transformation pairs (input/output sequences) from the crawled commits, we follow the following process. Within each commit, we parse the modified C\# files that end with the extension "*.cs" using the \hyperlink{https://github.com/tree-sitter/tree-sitter-c-sharp}{tree-sitter} parser. We first extract the classes within the file and its corresponding member methods. We apply some pre-processing steps on the method bodies by normalizing white-space and removing comments. This allows us to ignore any trivial modifications to whitespace and comments. Using the method signature, we identify the corresponding versions of the method in the before and after files. We then compare the normalized method bodies between the two versions of the file and discards the methods whose bodies do not appear to be have been modified. From here on, we refer to the remaining modified methods as focal methods.



\begin{figure*}
    \label{simplified_tree}
    \pgfplotsset{compat=newest}
    \centering
    {
        \begin{tikzpicture}[>=stealth', level/.style={level distance = 1cm}, scale=0.75, every node/.append style={transform shape}];
    \node
    {
    \lstset{linewidth=12cm,xrightmargin=.2\textwidth,basicstyle=\footnotesize\ttfamily}
    \begin{lstlisting}[frame=tlrb]
using System;
using System.Collections.Generic;
using System.Linq;
using Shared;
using Shared.Data.Extractors;
using Shared.Helpers;
using WindowsActivityTracker.Data;
using WindowsActivityTracker.Models;
class DayFragmentationTimeline : BaseVisualization, IVisualization
{
    /* begin */
    public override string GetHtml()
    {
        var html = string.Empty;
        var orderedTimelineList = Queries.GetDayTimelineData();
        var sum = orderedTimelineList.Sum(i => i.DurationInSeconds);
        if (orderedTimelineList.Count <= 3 || sum < 5*60)
        {
            html += VisHelper.NotEnoughData(Dict.NotEnoughData);
            return html;
        }

        html += GetActivityVisualization(orderedTimelineList);
        return html;
    }
    /* end */

    private string GetActivityVisualization(List<WindowsActivity> list)
    {
        var categories = list.Select(a => a.ActivityCategory).ToList();
        var html = string.Empty;
        html += @"<script type='text/javascript'>
                var onLoad=window.onload;
                window.onload=function()
                {
                    if(typeof onLoad=='function')
                    {
                        onLoad();
                    }
                }";
        html += "};";
        html += "</script>";
        html += "<div style='align:center'><p id='hover'>" +\
                defaultHoverText +\
                "</p></div>";
        html += "<div id='" +\
                activityTimeline +\ 
                "'align='center'></div>";
        html += LegendHelper.GetLegendForCategories(categories);
        return html;
    }

    private string defaultHoverText;
    private string activityTimeline;

    private string ReadableProcesses(List<WindowProcessItem> list);
    private string FormatWindowTitle(string windowTitle);
    private string FormatContextCategory(ActivityCategory category);
    private string FormatCategories(List<ActivityCategory> list);
}



    \end{lstlisting}
    };
    \node [text width=10cm, anchor=west, right,fill =white] at (7.2,0.2)
    {
    \lstset{linewidth=10cm,basicstyle=\footnotesize\ttfamily}
    \begin{lstlisting}[frame=tlrb]
using System.Text;
class DayFragmentationTimeline : BaseVisualization, IVisualization
{
    public override string GetHtml()
    {
        var html = string.Empty;
        var orderedTimelineList = Queries.GetDayTimelineData();
        var sum = orderedTimelineList.Sum(i => i.DurationInSeconds);
        if (orderedTimelineList.Count <= 3 || sum < 5*60)
        {
            html += VisHelper.NotEnoughData(Dict.NotEnoughData);
            return html;
        }

        _sb.Clear();
        GetActivityVisualization(orderedTimelineList, _sb);
        html += _sb.ToString();
        return html;
    }

    private void GetActivityVisualization(List<WindowsActivity> list,
                                                        StringBuilder sb)
    {
        var categories = list.Select(a => a.ActivityCategory).ToList();
        sb.Append(@"<script type='text/javascript'>
                  var onLoad=window.onload;
                  window.onload=function()
                  {
                      if(typeof onLoad=='function')
                      {
                          onLoad();
                      }
                  }");
        sb.Append("};");
        sb.Append("</script>");
        sb.Append("<div style='align:center'><p id='hover'>")
            .Append(defaultHoverText)
            .Append("</p></div>");
        sb.Append("<div id='")
            .Append(activityTimeline)
            .Append("'align='center'></div>");
        sb.Append(LegendHelper.GetLegendForCategories(categories));
    }

    private StringBuilder _sb = new StringBuilder();
}

    \end{lstlisting}
    };
    
    \draw [decorate,decoration={brace,amplitude=6pt,mirror,raise=4pt},yshift=0pt]
(0.,6.2) -- (0.,8.3) node [black,midway,xshift=1.42cm] {\scriptsize
Using Statements};

    \draw [decorate,decoration={brace,amplitude=10pt,mirror,raise=4pt},yshift=0pt]
(4.3,1.4) -- (4.3,5.1) node [black,midway,xshift=1.4cm] {\scriptsize
Focal Method};

    \draw [decorate,decoration={brace,amplitude=8pt,mirror,raise=4pt},yshift=0pt]
(5.,-5.6) -- (5.,1.) node [black,midway,xshift=1.28cm] {\scriptsize
Caller-Callee};

\draw [decorate,decoration={brace,amplitude=2.5pt,mirror,raise=4pt},yshift=0pt]
(-0.,-6.75) -- (-0.,-6.15) node [black,midway,xshift=1.2cm] {\scriptsize
Class Attributes};

\draw [decorate,decoration={brace,amplitude=6pt,mirror,raise=4pt},yshift=0pt]
(4.5,-8.3) -- (4.5,-6.9) node [black,midway,xshift=1.4cm] {\scriptsize
Method Signatures};

    \end{tikzpicture}
    }
  \caption{An example of an Input-output pair used in finetuning. The input consists of the focal method along with various class/file level context elements such as using statements within the file, class member attributes, focal method's caller/callee methods, other method signatures. We also include C-style comments (\textcolor{forestgreen}{\texttt{/* begin */}} and \textcolor{forestgreen}{\texttt{/* end */}}) before and after the focal method, indicating its location to the model. The example shown above comes from an actual performance commit made by a C\# developer to an open source repo and shows a perf change to replace a sequence of string concatenations with a \texttt{StringBuilder}. Such a change would save allocations as each string concatenation leads to a new string allocation, whereas the \texttt{StringBuilder} defers the allocation until after all the component strings are gathered and a call to \texttt{ToString()} is made. Additionally, the change also caches and re-uses the \texttt{StringBuilder} instance, as opposed to allocating a new one each time. In this case, the output also consists of an additional using statement, importing the namespace containing the definition of \texttt{StringBuilder}, along with modified versions of the focal and callee method and the new class attribute.}
\label{inputoutputpair}
\end{figure*}

Next we construct an input/output pair for each focal method. Figure \ref{inputoutputpair} shows an example of one such input/output pair. We start by including the focal method itself in the example input, whose location is indicated to the model using C-style comments (\texttt{/* edit */} and \texttt{/* end */}) before and after the focal method. Performance changes often require changes beyond the focal method itself (as seen in the second example in Figure \ref{motivating_example}), such as adding new class level attributes or additional imports, or even changes to other methods within the class that make calls to or are called by the focal method. 
Prior work has shown that including additional class/file-level context information with the focal method results in a higher quality predictions in such code generation tasks ~\cite{Tufano2020UnitTC, Drain2021DeepDebugFP}. We believe that such contextual information would prove useful in generating performance patches as well. Below are the file/class level context elements we include as part of the input:
\begin{itemize}
    \item \emph{Using Statements:} These tell the model what import statements exist within the file and whether new imports need to be added for the new methods or APIs used in the recommended changes.
    \item \emph{Class Attributes:} These are the containing class's member attributes. The underlying types of class attributes can often be important information in determining the right performance fix. This may include fixing incorrect usages of variables of certain types that cause performance issues or recommending a more appropriate data structure e.g. replacing \texttt{List<T>} with a \texttt{HashSet<T>}, etc.
    \item \emph{Caller-Callee Methods:} These are the methods that directly make calls to or are, in turn, called by the focal method. This information can help the model learn patterns of changes that involve hoisting/memoizing calls across methods, to optimize computations or simply modifying the caller/callee to be consistent with the modified focal method.
    \item \emph{Method Signatures:} Finally, we add the signatures for any other methods that aren't caller or callee methods of the focal method. Due to limited token space, we are unable to add the bodies of each method. This information could shed some light on the nature of the class itself and provide context as to what other methods are present in the class for the model to use in the generated patch.
\end{itemize}

Due to the input token window for BART-large being limited to only 1024 tokens, we construct the example input in an iterative fashion. We start by including the focal method in the example input and then incrementally add each contextual element in the list above. Before adding each type of contextual elements, we ensure that the resulting sequence will be within the allowed range of tokens i.e. $\leq$1024 tokens. This way, we try to incorporate as much context into the limited span of tokens while staying within the allowed limit. 


For the output, in addition to changes to the focal method, we include any of focal method's caller-callee methods that are modified by the commit. We also include any additional imports that may have been added as well as class attributes defined/modified that are used by the focal method or modified caller-callee methods. This way we allow the model to output patches that make changes to not only the focal method but also the caller-callee methods, class attributes as well as add any new methods, attributes and import statements as needed. Figure \ref{inputoutputpair}, shows an example of an input output pair generated using the steps above.

\subsubsection{Data splitting and De-duping}
We split the finetuning data on the project level. We leave out two sets of test and validation repos, each containing 600 repos that are not included in either step's training data. We also dedupe the examples in each set as well as remove any near duplicates~\cite{allamanis2019deduplication} among them to ensure no overlap between train and test data.

\begin{table}[htbp]
    \centering
    \footnotesize
    \caption{Number of commits and examples in our training data for the All Commits and Perf Commits finetuning steps.}
    \label{perf_non_perf}
    \label{table:exp1}
    \begin{tabular}{l l l l l}
        \textbf{Commit Type Data} & \textbf{\# of Commits} & \textbf{\# of Examples}  \\\hline\hline
        All Commits &  11M & 16M\\\hline
        Perf Commits &  535k & 1.5M\\
    \end{tabular}
\end{table}




\subsection{Finetuning}
\label{finetuning}
We finetune the code pretrained BART-large model on the task of generating a performance improvements, given an input sequence containing the focal method and contextual elements (as explained in Section \ref{generating_examples}). We perform a two-stage finetuning. We first teach the model how to C\# developers make changes in general by first finetuning our pretrained transformer model on examples from all commits. We refer to the resulting model as DeepDev-C\#. We then perform a second finetune step over DeepDev-C\# model, using the set of code transformations examples extracted from performance commits, to teach it specifically how developers make performance optimization changes. We refer to the final model as DeepPERF. To better understand whether the first finetuning step has any significant impact on the results, we train a third model, finetuned directly on code transformations extracted from performance commits. We refer to this model as DeepDev-Perf-Commits. In our evaluation, we compare the three models and discuss possible reasons for differences in their performances.

\section{Empirical Evaluation}
In this section, we first explain our baselines and evaluation metrics. We then cover our quantitative and qualitative analysis methodology and results. 


\subsection{Baselines and Evaluation Metrics}
We compare DeepPERF model with the following two models:
\begin{itemize}

\item \textbf{DeepDev-C\#}, which was first pretrained on English and Source code, followed by finetuning on code transformations extracted from all C\# commits in our training data. 
\item \textbf{DeepDev-Perf-Commits}, which was first pretrained on English and Source code, followed by finetuning on code transformations extracted from only performance commits (commits with a performance-related keyword in the commit message) in our training data. 
\end{itemize}

In order to compare the models' performances we define the following metrics: 
\begin{itemize}
    \item \textbf{Verbatim Match \%}: We report the number of examples where one of the model's suggestions was found to match verbatim with the developer patch i.e. the ground truth output in the input/output pair.
    \item \textbf{Abstracted Match \%}: For our comparison to be independent of variable name matching, we replace variable names with generic names of the form VAR\_\{i\} (e.g. VAR\_0, VAR\_1, etc.), where "i" is determined based on the relative order in which variables are encountered when traversing the parse tree. We then compare the abstracted versions of the model suggestions with the similarly abstracted developer patch and report how many were found to match.
    \item \textbf{CodeBLEU}: We measure the CodeBLEU~\cite{Ren2020CodeBLEUAM} scores to gauge the similarity of model output with the actual developer patches. In addition to n-gram matching of BLEU, CodeBLEU also compares abstract syntax trees (AST) and data flow between two programs. Thus, it takes into account the syntactic as well as semantic code similarity. We use the hyperparameters that were shown to have the highest correlation to human scores in the study i.e. $\alpha, \beta, \gamma, \delta = 0.1, 0.1, 0.4, 0.4$.
\end{itemize}

Through this experiment, we intend to answer the following research questions:
\begin{itemize}
    \item \textbf{RQ1}: Are both finetuning steps (All Commits and Perf Commits step) in our two-step finetuning necessary?
    \item \textbf{RQ2}: Is DeepPERF able to provide a wide-range of performance optimizations?
\end{itemize}

\newcommand{\specialcell}[2][c]{
  \begin{tabular}[#1]{@{}l@{}}#2\end{tabular}}

\begin{table*}[ht]
\footnotesize
\caption{\label{tab:dataset_summary} Three categories of performance issues in expert-verified dataset of performance improvements.}
\bigskip
\centering
\renewcommand{\arraystretch}{1.5}
\label{dataset_summary}
\resizebox{\textwidth}{!}{
 \begin{tabular}{p{0.2\linewidth}| p{0.72\linewidth}| p{0.1\linewidth}} 
 \hline
  Change Category & Some Examples of Kinds of Performance Optimizations Within Category & \# of Examples \\ [0.5ex] 
 \hline\hline
 High Level Change (C1) & \specialcell{Memoize results using \texttt{Dictionary}/\texttt{ConcurrentDictionary},\\
 Hoist computation/allocation to outer scope (loop, method, class, etc.),\\
 Cache and re-use types like \texttt{List}, \texttt{Dictionary}, \texttt{StringBuilder}, etc.\\
 Introduce fast-path to avoid unnecessary computation, etc. }& 29 \\ 
 \hline
 Suggest Different API/Data Structure  (C2) & \specialcell{Replace a series of string concatenations with a \texttt{StringBuilder},\\
 Use more suitable data structure (e.g. \texttt{List<T>.Contains()} $\rightarrow$ \texttt{HashSet<T>.Contains()}),\\
 Remove LINQ usage (e.g. \texttt{List<T>.Any()/Count()} $\rightarrow$ \texttt{List<T>.Count},\\
 \quad unroll query to explicit \texttt{for}/\texttt{foreach}-loop etc.), etc.} & 71\\
 \hline
 Improve Existing API/Data Structure Usage  (C3) & \specialcell{Condense/optimize LINQ queries (e.g. \texttt{Count() == 0} $\rightarrow$ \texttt{!Any()},\\       \quad\texttt{Where(<lambda>).FirstOrDefault()} $\rightarrow$ \texttt{FirstOrDefault(<lambda>)}, etc.),\\
 Use more appropriate method in API or fix API usage (e.g. initialize \texttt{List}/\texttt{Dictionary} with size, when known beforehand,\\
 \quad remove unnecessary calls to \texttt{ToList()}/\texttt{ToArray()} if enumerable is enumerated once,\\
 \quad read directly to \texttt{MemoryStream}, rather than reading to buffer and then to stream, etc.), etc.\\} & 34 \\
 \hline
\end{tabular}}
\end{table*}

\begin{table}[htbp]
    \centering
    \scriptsize
    \caption{Summary of the results of our three models over dataset comprising of all the examples generated from commits that had a performance related keyword in the commit message, in a hold-out set of 50 test repos.}
    \label{table:all_commits_eval}
    \begin{tabular}{c| c| c| c}
        Model & Verbatim     & Abstracted & CodeBLEU\\
              &  Match \% &  Match \%  &         \\\hline
        DeepDev-C\#        & 14.2  & 16.5 & 69.2\\\hline
        DeepDev-Perf-Commits & 13.3 & 16.4 & 68.8\\\hline
        DeepPERF          & 15.6 & 19.1 & 69.3\\\hline
    \end{tabular}
\end{table}

\subsection{Quantitative Analysis}
For the purposes of our evaluation, we picked a random set of repos from our test set, none of which had previously been seen by our models. We then collected all performance commits (commits with a performance-related keyword in the commit message) that change only one ".cs" file to filter out squash merges that include a variety of changes across multiple files.   
This process yields $\sim$1500 commits, resulting in a total of $\sim$2100 examples. We use this set of examples for our evaluation.  

\subsubsection{\textbf{Two-step Finetuning Ablation}}
Table \ref{table:all_commits_eval} shows the results of our 3 models over this dataset using the metrics we defined earlier. We see that our DeepPERF model performs better than the other two models and is able to get $\sim$16\% of the examples in this dataset verbatim and $\sim$19\% verbatim when the variable names are abstracted away. The 3 models achieve very similar CodeBLEU scores, which is expected since, due to the nature of the task, most of the code will be shared between the model output and ground truth for any well-trained model.
The major difference between DeepPERF and the DeepDev-Perf-Commits is that DeepPERF is first finetuned on examples generated from all C\# commits, followed by a smaller finetuning step on examples generated using commits with a performance related keyword in commit title/description. On the other hand, DeepDev-Perf-Commits is finetuned directly on the smaller set of performance commit examples. DeepDev-C\# differs from the two as it's only finetuned on all C\# commits, but not directly on performance commits.

Comparing the results of the three, the reason both DeepPERF and DeepDev-C\# perform better than DeepDev-Perf-Commits could be because finetuning on all commits allows the models to learn better representations for code by seeing more examples of how changes are made by C\# developers, since the dataset for All Commits step is almost 10 times larger than the one containing only examples from perf commits. There is also a possibility that there may be some performance improvement changes that aren't explicitly annotated within the commit message. For example, developers may not always explain every change they make in their commits and squash multiple changes into a single commit, mentioning only the most important in the commit message. We found several examples of such commits in our training data, where the changes contained performance optimizations, but the commit message did not include any performance related discussion. While, we don't know pervasive this is and just how many such "phantom" performance changes exist outside of performance commits, their presence would imply that models trained on All Commit data would overall be seeing more performance improvement code transformations during training than one that's been directly trained on performance commits.


\subsection{Qualitative Analysis}
To better understand the different types of performance improvements DeepPERF can suggest, we performed a qualitative evaluation on a subset containing $\sim$125 performance commits from our test set in the previous section. This resulted in a set of 132 examples demonstrating a variety of performance fixes, each of which were verified with two C\# performance experts. 
Based on our understanding of performance changes in C\#, we observe that the changes fall into the following three broad categories of performance improvements:
\begin{itemize}
    \item (Category 1) High Level Changes: These consist of algorithmic changes that require modifications to the overall code structure. These changes could include hoisting calls/allocations to an outer scope, adding caching/memoization to avoid repeated computation, introducing a fast-path, etc. 
    \item (Category 2) Suggesting Different API/Data Structure: These are language/API specific changes to replace or remove an existing API or data structure usage in favor of a better alternative. These changes could include removing LINQ by unrolling queries into explicit loops, suggesting a different data structure (e.g. replace List with a HashSet, when performing look-ups), etc.
    \item (Category 3) Improving Existing API/Data Structure Usage: These are also language/API specific changes, but suggest modifications to existing usage of an API or data structure when deemed incorrect or sub-optimal. These may include changes like condensing LINQ queries to be more optimal, fixing incorrect uses of a data structure, using a better suited overload of a library function, etc.
\end{itemize}

Table \ref{dataset_summary} shows some example performance changes found within the 3 categories as well as the number of examples in our dataset that fell within that category. We found that majority of these changes required deep knowledge of APIs/data structures or involved high level algorithmic modifications, which is not possible for the compiler to make automatically. We expect some analyzers to be able to fix a small portion of issues that fall in the second or third categories, but even these examples we found to be were quite varied. 

\subsubsection{\textbf{Human Evaluation Methodology}} For each example in our dataset, we sample 2000 hypotheses from the model and take the top 500 suggestions, based on the average likelihood of tokens. Since we have so many suggestions, 500 suggestions for $>$100 examples, it would be difficult to evaluate each of them by hand even with a team of experts. Therefore, we use the following evaluation metric to help us approximate the model's Top-K Accuracy. We report this in addition to the metrics mentioned earlier (i.e. Verbatim Match, Abstracted Match and CodeBLEU):
\begin{itemize}
    \item \textbf{Closest Match Top-K Accuracy \%}: Using a code search technique such as Aroma~\cite{luan2019aroma}, we find the document within the corpus of model suggestions that is closest to the developer patch, for each example. We then verify the most similar suggestion with two performance experts, neither of whom are on the author list. The experts are shown both the developer change and model suggestion and asked to assess whether they considered the model suggestion to be making the same performance optimization and are semantically the same as the developer change.
\end{itemize}

\begin{table}[htbp]
    \centering
    \tiny
    \caption{Summary of the results of our three models over the manually curated dataset.}
    \label{table:collected_data_eval}
    \begin{tabular}{c| c c c c| c| c| c}
        & \multicolumn{4}{c|}{Closest Match} & Verbatim & Abstracted & \\
         Model &  \multicolumn{4}{c|}{Top-K Accuracy \%} & Match \% & Match \% & CodeBLEU  \\
         &  1 & 10 & 100 & 500 & & &  \\\hline
        DeepDev-C\# & 2.3  & 14.4 & 31.1 & 37.9 & 24.6 & 26.1 & 68.3\\\hline
        DeepDev-Perf-Commits & 7.6 & 18.9 & 31.2 & 42.4 & 26.1 & 29.1 & 70.6\\\hline
        DeepPERF &  8.3 & 18.2  & 34.1 & 53.0 & 34.3 & 37.3 & 70.7\\\hline
    \end{tabular}
\end{table}

\begin{figure}[h]
\centering
\includegraphics[width=0.48\textwidth]{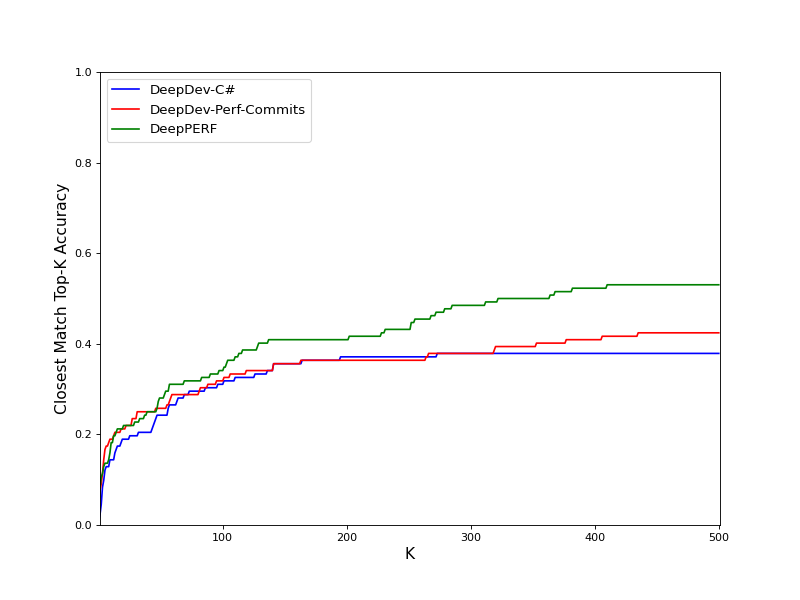}
\caption{Closest Match Top-K Accuracy plot of our models on the manually curated dataset of performance changes. We can see that DeepPERF achieves the best score among the 3 models.}
\label{accuracy_plot}
\end{figure}

\begin{figure}
  \begin{tikzpicture}
  \centering
  \begin{axis}[
        ybar, axis on top,
        title={},
        height=8cm, width=8cm,
        bar width=0.3cm,
        ymajorgrids, tick align=inside,
        enlarge y limits={value=.1,upper},
        enlarge x limits={abs=25pt},
        ymin=0,
        ymax=100,
        axis x line*=bottom,
        axis y line*=left,
        tickwidth=1pt,
        legend cell align={left},
        legend style={
            nodes={scale=0.7, transform shape},
            anchor=east,
            legend columns=1
        },
        ylabel={Percentage (\%)},
        xlabel={Performance Improvement Category},
        symbolic x coords={
           Category 1,Category 2,Category 3},
       xtick=data,
       nodes near coords={
        \pgfmathprintnumber[precision=0]{\pgfplotspointmeta}
       }
    ]
    \addplot [draw=none, fill=blue!40] coordinates {
      (Category 1,24.1) 
      (Category 2,32.4)
      (Category 3,58.8) };
   \addplot [draw=none,fill=red!40] coordinates {
      (Category 1,27.6) 
      (Category 2,40.8)
      (Category 3,55.9) };
   \addplot [draw=none, fill=green!40] coordinates {
      (Category 1,37.9) 
      (Category 2,49.3)
      (Category 3,67.6) };

    \legend{DeepDev-C\#,DeepDev-Perf-Commits,DeepPERF}
  \end{axis}
  \end{tikzpicture}
  \caption{Performance of our models on the three categories of performance issues: High Level Changes (Category 1), Suggesting Different API/Data Structure (Category 2), Improving Existing API/Data Structure Usage (Category 3). We can see that the DeepPERF model (green) tends to perform the best among the models in all three categories, followed by DeepDev-Perf-Commits and DeepDev-C\#.}
\label{category_plot}
\end{figure}
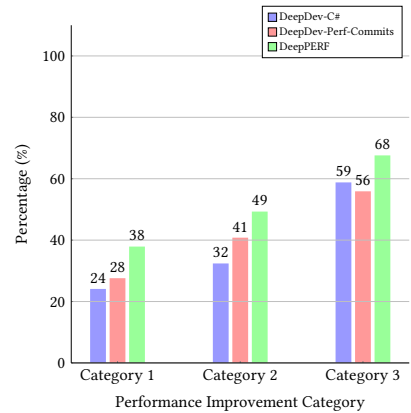

\subsubsection{\textbf{Results}} Table \ref{table:collected_data_eval} and Figure \ref{accuracy_plot}, show the results of the 3 models using our defined metrics. We see that our best model is able to solve $\sim$53\% of the examples in our dataset, getting $\sim$34\% verbatim as the developer fix. The Closest Match Top-K Accuracy plot was computed based on the associated rank of the retrieved suggestion among the top-500 hypotheses, when found to be correct by both of the two C\# performance experts. In majority of the cases, the main reasons for dissimilarities from the developer were the model suggesting different variable names, or other slight variations like using the \texttt{var} keyword instead of the variable's type or using a \texttt{for}-loop as opposed to a \texttt{foreach}-loop where both are appropriate, difference in order of statements where relative order did not matter (such as \texttt{using} statements at the start of file), etc. Figure \ref{category_plot} shows the performance of our models in the 3 categories of performance changes. We also see that our best model out-performs the other two models in each category.

\vspace*{0.25cm}
\begin{rqbox}
\textbf{RQ1 \& RQ2}: In summary, we conclude that both finetuning steps were necessary as DeepPERF clearly demonstrates better performance than the other two models on the overall test dataset of performance commits as well as the smaller expert verified dataset of performance optimizations. Additionally, we see that DeepPERF can generate suggestions that span a wide variety of performance optimizations encompassing both high-level algorithmic to low-level API/Data structure related performance changes. Furthermore, these changes were considered equivalent to the developer-made performance improvements by two performance experts in C\#.
\end{rqbox}

\begin{figure}[h]
\centering
\includegraphics[width=0.48\textwidth]{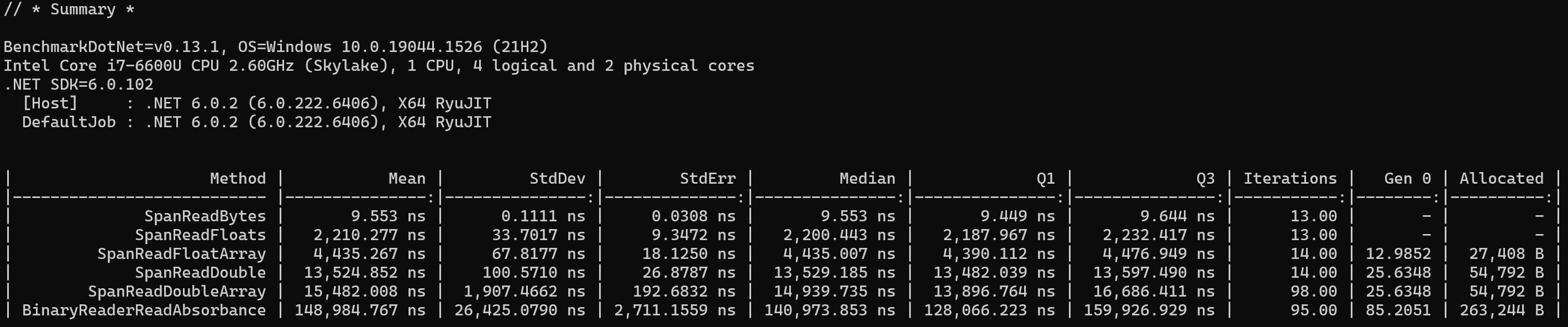}
\caption{An example of the output generated from executing a BenchmarkDotNet test suite. 
The first column shows the different benchmarks defined by the user. BenchmarkDotNet automatically runs each of these user-defined benchmarks multiple times and reports metrics such as sample mean, standard deviation, standard error, median, first and third quartile of the test duration. For allocations, it reports the memory allocated on average during each test run.} 
\label{benchmarksummary}
\end{figure}

\section{In-the-wild Evaluation}
To see whether our approach can suggest real performance improvements to existing C\# projects, we performed an "in-the-wild" evaluation of DeepPERF on a set of 50 C\# GitHub repos, not previously seen by our model. For this evaluation, we chose repos that contained both benchmark and unit tests to allow us to verify that our suggested changes are correct (using unit tests) and lead to measurable performance improvements (using benchmark tests). 
Benchmark tests in C\# are usually written using the \hyperlink{https://github.com/dotnet/BenchmarkDotNet}{BenchmarkDotNet} library. Shown in Figure \ref{benchmarksummary} is an output summary of a BenchmarkDotNet test. BenchmarkDotNet automatically runs each benchmark test in the user-written test suite multiple times and reports metrics for the duration of a given benchmark test as well as the amount of memory that is allocated on average on each run. It can also give other information such as how frequently Generational GC is triggered.

Through this experiment, we intend to address the following research questions:
\begin{itemize}
    \item \textbf{RQ3}: What are the reasons behind some of our model's changes failing to compile? How could this be improved?
    \item \textbf{RQ4}: Is DeepPERF able to suggest changes that lead to real performance improvements? If so, how much performance improvement do these changes typically provide? Are these suggestions considered useful by the developers?
    \item \textbf{RQ5}: How effective are unit and benchmark tests in ensuring changes are correct performance improvements?
\end{itemize}
\subsection{Experiment Setup}

To find repos with Benchmark tests, we sampled 50 C\# repos where BenchmarkDotNet NuGet package is mentioned in one of the build configuration files ("*.csproj") in the repo. To limit the methods we need to generate changes for, we select methods that satisfy following two criteria: (1) are present on the execution path of the repo's benchmark tests, (2) have a high line/branch coverage with the repo's unit tests. This yields 201 methods across the 50 test repos. We then generate model inputs for each of these methods, including contextual information as described in Example generation step (Sec. \ref{generating_examples}). We use our best model, DeepPERF, and sample 2000 suggestions for each of these inputs. We then pick the top 100 suggestions based on their average token likelihood, and test each of these suggestions against the repo's main branch. 

From these suggestions, we first filter out changes that are syntactically incorrect. We found that $\sim$7\% of the suggestions had a syntax error. Most of there were due to early truncation or repetition when generating long outputs, which are known issues when generating text using such language models.

\subsection{Running Unit Tests}
We then run unit tests for each of the remaining $\sim$93\% suggestions. This step filters out suggestions that fail to compile or are found incorrect based on the unit test cases provided by the developer. Table \ref{dynamic_evaluation} shows a breakdown of how many suggestions fail at this stage. As we can see, at the end of this step we are left with $\sim$44\% of the suggestions we started with. 

\begin{table}[htbp]
    \centering
    \scriptsize
    \caption{Breakdown of the results of running unit tests.}
    \label{dynamic_evaluation}
    \label{table:exp1}
    \begin{tabular}{l l l}
        \textbf{Result} & \textbf{Occurrences} & \textbf{\% of Suggestions} \\\hline\hline
        Syntax Error &  1329 & 6.6\\\hline
        Compilation Error &  7860 & 39.1\\\hline
        Failed Unit Tests &  2056 & 10.2\\\hline
        Passed Unit Tests &  8855 & 44.1\\\hline\hline
        Total &  20100 = 201 * 100 & 100\%\\
    \end{tabular}
\end{table}


\subsection{Analyzing Build Failures (RQ3)}
Table \ref{compilation_errors} shows the main reasons of compilation errors. After grouping together the first compilation error in each suggestion that fails to compile, we found that they fell into 4 major error categories: \emph{Undefined Identifier}, \emph{Incorrect Argument passing}, \emph{Incorrect Using Statements} and \emph{Incorrect Return Type}. Upon looking at some instances of each category, we identified patterns of mistakes in the model's suggestions that cause these errors. 

We noticed that the \emph{Undefined Identifier} errors tend to happen when the model tries to use methods or classes outside provided context. As the model can only guess what other classes exist in the project and the methods contained within, it sometimes makes calls to methods that do not exist. We believe this could be improved by incorporating additional information regarding other classes within the project to the input, such as the classes being used in the focal method or within imported namespaces. 

The \emph{Incorrect Argument} errors also tend to occur when the model calls a method outside of provided the context. This results in the model passing in the wrong arguments types or number of arguments by making calls to method overloads that don't exist. We often saw this occur when the model tried to call member methods within some project-specific classes that were instantiated somewhere in the input code.

Cases for the \emph{Incorrect Using Statements} follow a similar pattern as well. Here the model tries to import namespaces within the repo that don't exist or from packages that aren't in the build files. Since it doesn't know what other files exist in the project or the packages included in build, it often adds incorrect import statements. 

The fourth category, \emph{Type Mismatch}, occurs when the model suggests modifications that change the types of one or more class attributes, which get used elsewhere in the class. Since it can only modify the methods that are included in the input context (due to limited window), it is unable to modify these other methods. Other reasons for these errors include mismatch caused by changing the return type of a method, when the input class implements an interface, since changing the type would cause the method in the parent to not be overridden, leading to a compiler error.

Based on these observations, we believe a significant portion of above errors could be resolved by including a larger context containing more methods in the input class or even other classes/files in the project, through extended context~\cite{ewashemnlp}. We leave this exploration to future work.


\begin{table}[htbp]
    \centering
    \scriptsize
    \caption{Main reasons for compilation errors.}
    \label{compilation_errors}
    \begin{tabular}{l l l l c}
        \textbf{Error Cause} & \textbf{Error Codes} & \textbf{Occurrences} & \textbf{\% of} \\
        & & & \textbf{Errors}\\\hline\hline
        Undefined Identifier & CS1061, CS0117, & 3672 & 46.7 \\
        & CS0246, CS0103, & &\\
        & CS1579, etc. & & \\\hline
        Incorrect Arguments & CS1503, CS1501, & 2758 & 35.1 \\
         & CS1729, CS7036, & & \\
         & CS0305, CS0029, & &\\
         & CS0019, etc. & &\\\hline
         Incorrect Using Statements & CS0234 & 610 & 7.8 \\\hline
        Type Mismatch & CS0266, CS0738, & 246 & 3.1 \\
        & CS0508, etc. & & \\\hline
        Other Mistakes & CS0021,CS0122, & & \\
        & ($\sim$120 misc. codes) & 574 & 7.3 \\\hline\hline
        Total &  & 7860 & 100.0 \\
        
    \end{tabular}
\end{table}

\subsection{Running Benchmark Tests}
The next step is to run benchmark tests for each of the changes that pass unit testing stage. However, before we run the benchmark tests we had to make some changes to the provided benchmark test suite to ensure the tests track the right metrics and that results are comparable among separate runs. By default, BenchmarkDotNet tests do not track allocations. For 22 out of the test repos, we found that memory tracking wasn't enabled and we had to enable it ourselves by adding a \mintinline{CSharp}{[MemoryDiagnoser]} attribute to the class containing the benchmarks. Changing this does not affect the results for other metrics tracked by the benchmarks like test durations. Another change we had to make to make the numbers comparable between separate runs was to add seeds to instances of random number generators instantiated in the benchmarking code. This is to ensure that the tests are deterministic so that the results can be compared between separate runs of the tests. 

Additionally, to ensure no interference from background processes, we run the benchmark test in a sterile work environment with minimal workload other than the test itself. We first run the benchmark tests without any changes to measure the baseline performance of the application and then once after applying each of the changes that passed unit testing.

\subsection{Analyzing Benchmark Results (RQ4)}
\subsubsection{Comparing Against Baseline}
Allocations are expected to stay consistent for C\# applications as long as the benchmark tests are deterministic, so it is easy to tell if the change has improved memory usage by comparing the "Allocated" column (as shown in Figure \ref{benchmarksummary}). We consider a change to be a performance improvement in terms of Memory if it reduces allocations compared to the baseline. 

For test duration, we use the "Mean", "StdDev" and "Iterations" columns, representing the sample mean, standard deviation and size, respectively. We make the assumption that the test duration readings are normally distributed. For each benchmark in the suggestion sample, we conduct a one-tailed Welch's \emph{t}-test at 5\% significance level to determine if the population mean of the suggestion code's sample is less than the population mean of the baseline (unmodified code) sample for the corresponding benchmark. In other words, our null hypothesis is that the population means of two samples are equal and the alternative hypothesis is that the population mean for suggestion is lower than baseline. We discard the suggestions that fail to reject the null-hypothesis. Following this, we conduct some additional checks on the remaining suggestions to reduce false positives and to ensure the change provides a significant enough improvement to be reported to the user. For this check, we use the "Q1" and "Q3" columns, which represent the first and third quartiles of the sample, respectively. We consider a suggestion to be a significant performance improvement over the baseline in terms of test durations, if the suggestion's upper Tukey fence is found to be lower than the the baseline's lower Tukey fence i.e. if $Q_{3_{suggestion}} + 1.5(IQR_{suggestion}) < (Q_{1_{baseline}} - 1.5(IQR_{baseline})$, where IQR is the interquartile range, $Q_{3} - Q_{1}$. Since there may be noise from background processes, this criteria also allows us to be robust to outliers and have fewer false positives. Finally, we also ensure that an improvement in allocation or test duration does not cause the other to deteriorate.

\subsubsection{Submitting a Perf Improvement PR}
Upon comparing the results against the baseline, we found that 543 suggestions improve performance metrics. 
These changes were saturated within 41 of the 201 methods. For each method, we verify up to 10 suggestions that pass unit tests and improve memory/test duration with a performance expert and submit a PR containing the first change that is a valid improvement. In case a project has correct suggestions for multiple methods, we squash all changes into a single PR.

For the cases where the model had generated correct suggestions, it was usually able to suggest the correct patch within the first or the second suggestion that passed unit tests and improved performance. Often times it suggested multiple distinct correct patches that seemed to improve performance. Figure \ref{motivating_example} shows two examples of valid performance improvement patches suggested during this evaluation that have been approved by the project owners.

\begin{figure}[h]
\centering
\includegraphics[width=0.4\textwidth]{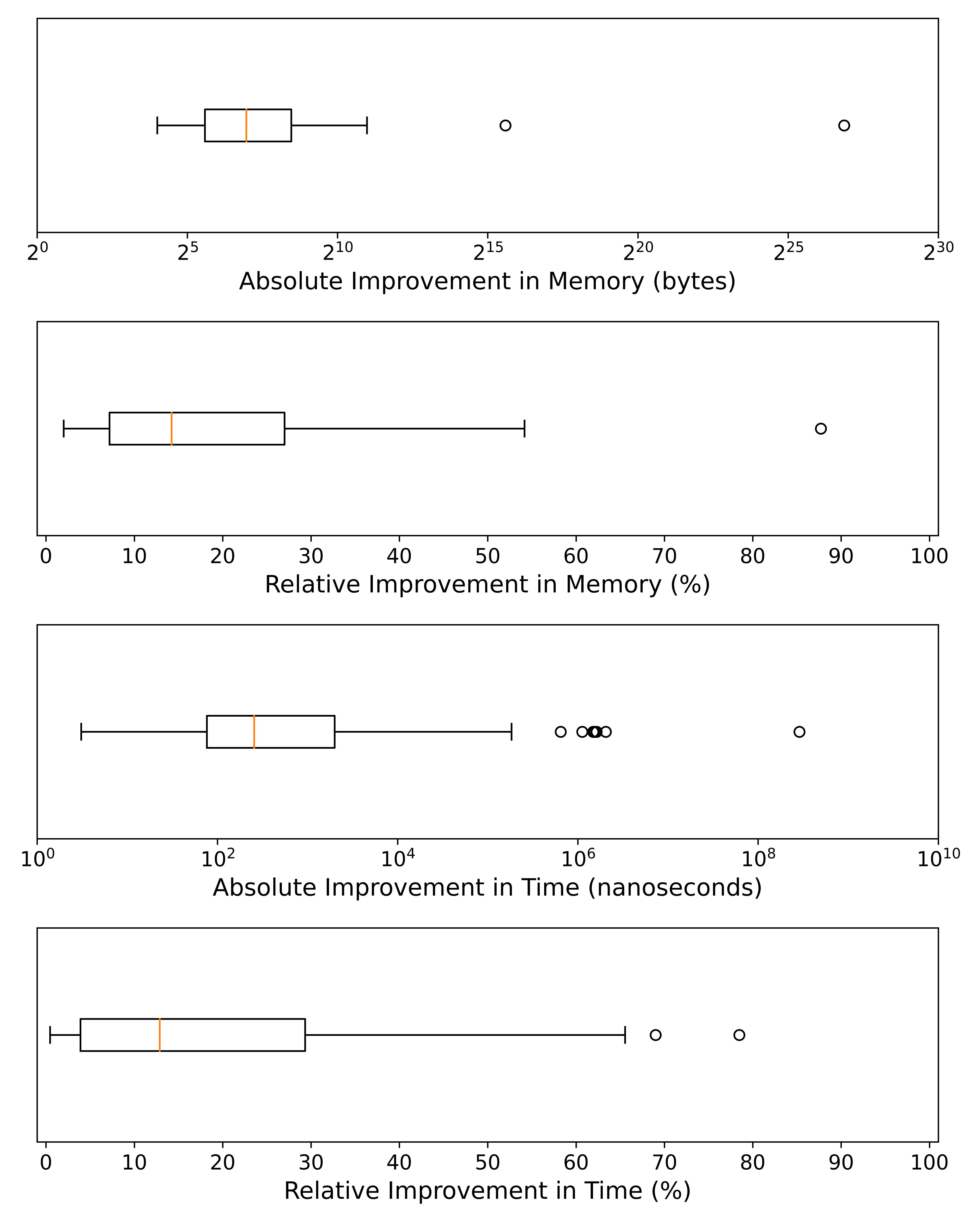}
\caption{Above boxplots show improvement in benchmark test durations and allocations over baseline due to DeepPERF suggestions. Top two boxplots show the absolute and relative improvement over baseline allocations, respectively. Similarly, the bottom two show the absolute and relative improvement over baseline test durations.}
\label{allocation_improvement}
\end{figure}

Figure \ref{allocation_improvement} shows the improvement in benchmark durations and allocations due to DeepPERF's suggestions compared to baseline numbers. Looking at the relative improvement, we see that DeepPERF's suggestions typically provide 10-15\% improvement in terms of both allocations and test duration. Interestingly, for allocations, a few of our suggestions provide improvement on the order of KBs or even MBs. While others on the lower end do seem to veer in the territory of micro-optimizations, it should be noted that this is from benchmark tests and the developer may have simply written their benchmark test to be small (i.e. fewer iterations). We also don't know how often the tested code is run when the application sees use by a real customer. Depending on how often the code being tested is exercised during the application's runtime, especially if it appears on the application's hot-path, even these smaller improvements could improve performance significantly. 

\vspace*{0.25cm}
\begin{rqbox}
\textbf{RQ4}: In summary, we found that for 28 out of the 41 methods, DeepPERF had at least one correct performance improvement suggestion. These changes usually provided a 10-15\% improvement over baseline in terms of memory or test times. We've submitted a total of 19 PRs, 11 of which have since been approved by the project owners demonstrating the usefulness of our suggestions. 
\end{rqbox}

\subsection{\textbf{False Positives (RQ5)}}
One of our PRs was closed because the repo was not open to external contributions. However, the developer did not comment as to whether they considered the changes to be incorrect. Our remaining 8 PRs are still "Open" waiting for a response from the project owner.
13 out of the 41 methods DeepPERF found an optimization for turned out to be false positives i.e. they only had incorrect suggestions that seemed to improve benchmark results and somehow managed passed unit tests. This is a known issue in such models as they often generate suggestions that are test suite adequate, but otherwise turn out to be incorrect. While we make sure the methods we test have a high code coverage, that doesn't guarantee that the unit test will detect all mistakes as it may not be written to specifically test the particular method being modified. Another reason could be that the test suite itself is lacking. One way to combat these cases would be to generate additional unit cases and use them as further validation in addition to user-provided unit tests. One could also train an additional classifier to determine whether a change is correct and use it for filtration. We leave these explorations for future work.

\vspace*{0.25cm}
\begin{rqbox}
\textbf{RQ5}: For a majority of methods, $\sim$68\%, that DeepPERF found improvements for, the changes were found to be valid and correctly passed the unit tests. However, in 13 out of 41 methods we only found invalid suggestions some of which were able to pass unit tests. While this is not insignificant, we believe this can be addressed by the means of generating additional unit tests or training a classifier to identify such cases. We leave these explorations to future work. 
\end{rqbox}





\section{Threats to Validity}
DeepPERF focuses on single file performance improvements, but often performance changes require modifications to multiple classes or even files. To generalize our approach to multiple file performance improvements, one could build on ideas like extended context~\cite{ewashemnlp} and extending the transformer's input embedding matrices~\cite{Drain2021DeepDebugFP} to be able to pass in a larger context potentially from multiple classes or files. Another challenge when constructing input/output pairs from a given commit, would be determining which changes within the commit are related. A possible way to address this could be to by establishing caller-callee information between methods across files or use import statements to see which files are likely connected to the change. At inference time, one could generate the input based on caller-callee relationship among files and apply suggested changes to all the files involved.

Our "in-the-wild" evaluation used benchmark tests to validate the performance gains from our suggestions. It is difficult to know how much, if any, performance improvement this would provide to the end-user of the application. But, the fact that the developer wrote benchmark tests for these methods is a strong indication that the code must be frequently exercised and expected to be on the application's hot-path. Future work could replace benchmarking and combine our model with profiling or load testing instead to assess the performance gains in a more realistic usage scenario.


\section{Related Work}
We describe how are work complements prior work in performance bug detection as well as automated bug detection. 

\subsection{Performance Bug Detection \& Fix} 
There is a rich history of building tools for detecting performance bugs and improving performance. The majority of these tools identify code locations that take a long time to execute. Several tools generate or select tests for performance testing \cite{zhang2011ASE, Grechanik2012ICSE, burnim2009ICSE}. Other performance detection tools focus on detecting a specific type of performance bug. For instance, a set of tools have been developed for detecting runtime bloat \cite{xu-pldi-2009, xu2010pldi1, dufour2008fse}, low-utility data structures \cite{xu2010pldi}, database related performance anti-patterns \cite{chen2014icse}, false sharing problem in multi-threaded software \cite{liu2011oopsal}, and detecting inefficient loops \cite{nistor2013, xiao2013issta}. Approaches fixing specific performance issues, such as repeated computations~\cite{memoization}, software misconfigurations~\cite{misconfigurations}, loop inefficiencies~\cite{caramelnistor}, etc. have also been developed.
Our tool extends the prior work on performance bug detection and fix by developing a system that focuses on alleviating general performance problems and considers both source code features as well as performance symptoms through benchmarking. 


\subsection{Automatic Bug Detection}
Prior work has investigated the use of static analyzers for detecting software bugs \cite{xu2010memory, viega2000its4, sonarqube, coverity}.
More recently, researchers have started to explore the usage of machine learning for both software bug detection and bug fix. For instance, in C/C++, VulDeePecker uses deep learning to detect two types of vulnerabilities. Similarly, Russell et al. \cite{russell2018automated} propose a machine learning based method vulnerability detection in C/C++ code bases. In Java, Pang et al. \cite{pang2015predicting} trained a machine learning model to predict static analyzer labels for Java source code. DeepFix leverages deep learning to generate fixes for simple syntax erros \cite{gupta2017deepfix}.  
We are uniquely contributing to this area of research by leveraging neural networks for detecting optimization opportunities and suggesting performance improvements.

\section{Conclusions}
Detecting and fixing performance bugs remains an important yet challenging problem in the software development process. Our work makes three contributes to address this problem. First, we present a novel transformer based model to automatically generate patches providing performance improvement. 
Second, we conduct an empirical evaluation of our model to show that it outperforms the baselines over a dataset of performance optimizations collected from performance commits made by C\# developers to open source repos on GitHub. Through this evaluation, we showed that our model is able to provide a wide-range of performance optimizations, which were verified by performance experts. 
Finally, we present a highly practical, end-to-end pipeline showcasing our vision for automatically generating performance improvements for real world projects. This pipeline consists of our model alongside unit-testing and benchmarking, which are used to validate the generated patches. We show that our model is able to suggest valid performance improvements that lead to tangible performance gains to real world applications. We submit pull-requests containing the optimizations generated by this pipeline. Several of these PRs have since been merged, showing that our changes are considered valuable by the project owners.


\bibliographystyle{ACM-Reference-Format}
\bibliography{references}
\end{document}